\documentclass[11pt]{article}
\usepackage{amsmath}
\usepackage{amsfonts}
\usepackage{amssymb}
\usepackage{graphicx}

\def\be\begin{Eq.uation}
 \def\ee{\end{equation}}
\def\bea{\begin{eqnarray}}
\def\eea{\end{eqnarray}}

\begin{document}
\begin{center}
\LARGE {Holographic Dark Energy Like in  $f(R)$ Gravity }
\end{center}
\begin{center}
{\bf $^{a,b} $ Kh. Saaidi}\footnote{ksaaidi@uok.ac.ir},
{\bf $^{b}$A.Aghamohammadi\footnote{ agha35484@yahoo.com}},

{\it $^a$Department of Physics, Faculty of Science, University of
Kurdistan,  Sanandaj, Iran}\\
{\it $^b$Faculty of Science,  Islamic Azad University of  Sanandaj, Sanandaj, Iran }
\end{center}
 \vskip 1cm
\begin{center}
{\bf{Abstract}}
\end{center}

We investigate the corresponding relation between  $f(R)$ gravity and holographic dark energy. We introduce a kind of  energy density  from $f(R)$ which has   role of  the same as    holographic dark energy.
 We obtain the differential equation that specify the evolution of the introduced energy density parameter based on varying gravitational constant. We find out a relation  for the equation of state parameter    to low redshifts which containing varying $G$ correction.
 \\

\newpage
\section{Introductions}
  Observational data\cite{1,2,3}, indicates  that there are several shortcoming of standard Gravity (SGR)\cite{6,7,8} related to cosmology, large scale structure and quantum field theory. Several attempts try to solve these problems by adding new and unjustified components in order to give an  adaptable picture for present universe. An approach for solving this problems, is to add new cosmic fluids in the right hand side of equations, which can cause a clustered structures (dark matter) or to accelerated dynamics (dark energy). Although, there isn't a quantum theory of gravity, one can proceed to investigate the nature of dark energy based on some principles of quantum gravity. A tremendous try in this regard is dubbed holographic dark energy (HDE) proposal \cite{10, 11}. As a rule, in the quantum field theory, $\rho_{\Lambda}$  as zero-point energy density is defined  based on $L$  (the  size of the current universe) as follow
 \begin{equation} \label{1e}
\rho_{\Lambda}= 3c^2 M_p^2L^{-2},
\end{equation}
where $c^2$ is a numerical constant of order unity and $M_p^{-2}=8\pi G$ is denoted the reduced planck mass. This $\rho_{\Lambda}$ is  comparable to the present  dark energy \cite{1c,2c}. The HDE model may be satisfy the coincidence problem, i.e why matter and dark energy densities are comparable, however, they have different equation of motion \cite{2c}. Lately, the holographic dark energy model has been extended and constrained by various astronomical observaations\cite{3c}.  It also has been extended to involve the spatial curvature  contribution \cite{4c}.

 An alternative approach is extended Einstein's theory of gravity, that $f(R)$ gravity is one of   the extended theories of gravity candidate,  which  itself was advocated to account for this accelerating universe \cite{8f,9f,10f,11f,12f,13f,14f,15f,16f,17f,n8,9n,11n,n10}. This kind of theories is fruitful and economics with respect to other theories. The assumption is that the Ricci scalar of the Einstein-Hilbert action is replaced by an arbitrary function $f(R)=R+h(R)$. The $f(R)$ gravity is assumed as replace dark energy.

In this work, we  want study an equivalence between $f(R)$ gravity and  Holographic dark energy. In this regards, we defined a dark energy density- like with transfer some terms of generalized Friedmann equation to rhs of equation field. As a matter of fact, equivalency between $f(R)$ gravity and dark energy like is as well as   connection between $\Lambda$ terms and dark energy. Note that,  at first, Einstein added  $\Lambda$ term to  the his theory as a geometrical term.  As a rule, in most the dark energy models investigated Newton's gravitation constant assume to be constant. Here, we  consider Newton's gravitation constant to be variable with time. There are noticeable reasons that the $G$ can vary with time or being a function of the scale factor \cite{31c}. Among, helio-seismological\cite{32c}, astro-seismological data from the pulsating white dwarf star G117-B15A \cite{33c}, Hulse-Taylor binary pulsar\cite{34c,35c}, result in $\mid\dot{G}/G\mid\lessapprox 4.10\times 10^{-11}yr^{-1}$ for $z\lesssim 3.5$\cite{36c}. In addition, a changing $G$ has some theoretical merit too, the discord in  Hubble parameter value\cite{37}, the cosmic coincidence  problem\cite{38}, allaying the dark matter problem \cite{39}.

This paper is organized as follows: in the next section we will a review of $f(R)$ gravity cosmology. In the third section, we found  the holographic dark energy like  with gravitation constant depend on time and derive the differential equation that specify the evolution of dark energy parameter. In the four section we obtain the parameter of the dark energy equation of state at the low redshift. Eventually, the latter section is devoted to conclusion.
\section{Theories of the $f(R)$ gravity}
We consider a class of modified gravity in which modifies Einstein- Hilbert action by replacing Ricci curvature scalar by an arbitrary function of curvature as follow:
\begin{equation}\label{2h}
S=\int \left(\frac{f(R)}{2}+kL_m \right)\sqrt{-g}d^4x,
\end{equation}
which is introduced in cosmological contexts and $k=8\pi G$. Variation with respect to metric yields   gravitational equations of motion as
\begin{equation}\label{3h}
R_{\mu\nu}f '-\frac{1}{2}fg_{\mu\nu}+\left( g_{\mu\nu}\Box-\nabla_{\mu}\nabla_{\nu}\right)f '=kT_{\mu\nu},
\end{equation}
where prime is the derivation with respect to curvature scalar (R) and $\Box$ is covariant  D'Alembert operator ($\Box\equiv \nabla_{\alpha}\nabla^{\alpha}$).
The equation of motion for a new scalar degree of freedom is given by the trace of equation (\ref{3h})
as \cite{9n}
\begin{eqnarray}\label{h4}
\Box f '=\frac{kT}{3}+\frac{\left[  2f-Rf '\right]   }{3}.
\end{eqnarray}
   We redefine  the scalar degree of freedom by
     \begin{eqnarray}\label{5h}
\phi=f '-1.
\end{eqnarray}
Therefore, the equation (\ref{h4}) in the form of equation of motion of canonical dimensionless scaler field $\phi$ with a force term ${\cal F}$ and potential $V$ is as
\begin{eqnarray}\label{6h}
\Box \phi= V'(\phi)-{\cal F},
\end{eqnarray}
where $V'(\phi)=\frac{1}{3}\left(  2f-Rf '\right) $, and the force term that drives the scalar field $\phi$ is a trace of the stress-energy tensor ${\cal F}=\frac{8\pi G}{3}T$.
Considering, a homogeneous cosmological model in $f(R)$ gravity with the usual matter field, the length element describing expansion of the universes is expressed by a flat Friedmann-Robertson-Walker metric as
   \begin{equation}\label{7h}
ds^2= dt^2-a^2(t)(dx^2 +dy^2+dz^2) ,
\end{equation}
using equations (\ref{6h},\ref{7h}), the scalar degree of freedom $\phi$ obey a usual scalar field equation with a force term on the right hand as
\begin{equation}\label{h10}
\ddot{\phi}+3H\dot{\phi}+V'(\phi)={\cal F}.
\end{equation}
Considering, $tt$ component of gravitational equations from equation (\ref{3h}) for the metric (\ref{7h}), we have
\begin{eqnarray}\label{8h}
3H\left(f '\right)^{.}-3\frac{\ddot{a}}{a}f '+\frac{1}{2}f=8\pi G \rho.
\end{eqnarray}
 Here, it is clear that analogy equation (\ref{8h}) in the $f(R)$ cosmology with Fredmann equation is not transparent.To get a proper limit, let us we write $\ddot{a}$ with respect to the curvature scalar by using
\begin{eqnarray}\label{9h}
R=6\left(\frac{\ddot{a}}{a}+\left( \frac{\dot{a}}{a}\right)^2  \right).
\end{eqnarray}
Finally, the equation (\ref{8h}) yield as follows
\begin{eqnarray}\label{10h}
H^2+\left(\ln f ' \right)^{.}+\frac{1}{6}\left( \frac{f-Rf '}{f '}\right)=\frac{8\pi G}{3f '}\rho.
\end{eqnarray}
In comparison with Friedmann equation, it is clear if $f '\rightarrow 1$ then equation (\ref{10h}) is left in  the usual Friedmann equation form. In the general case, the extra terms are functions of scalar degree of  freedom $\phi$ and its first time derivative. Hence, we can re-write the equation (\ref{10h}) in the following standard form
   \begin{eqnarray}\label{11h}
H^2=\frac{8\pi G }{3\left(1+\phi \right) }\left( \rho_m+\rho_f\right),
\end{eqnarray}
where, $\rho_f$ is assumed as dark energy density as $\rho_{\Lambda}$. We can define $\rho_f$ as   follows
\begin{equation}\label{12h}
\rho_f= -\frac{3\left(1+\phi \right)  }{8\pi G}\left( \left(\ln f ' \right)^{.}+\frac{1}{6}\left( \frac{f-Rf '}{f '}\right)\right)
\end{equation}
\section{Holographic Dark Energy Like  With Gravitational Constant Depend on Time }
\subsection{The Density Parameter}
We will to set up the holographic dark energy-like  for a Newton's constant (G) depend on time. Therefore, in this regard, we consider the metric given by the equation (\ref{7h}).
 Concerning, the form of the first Friedmann equation, from Eq (\ref{11h}) we have:
\begin{eqnarray}\label{2g}
H^2=\frac{8\pi G}{3\left( \phi+1\right) }\rho.
\end{eqnarray}
Here, $\rho=\rho_m+\rho_f$ is the energy density, $\rho_m=\rho_{m_0} a^{-3},\rho_f$ are dark matter density and dark energy density like  respectively and $\rho_{m_0}$ indicate the present value that quantity.    Then, from   the Eq.(\ref{11h}), we introduce the effective density parameter $\Omega_{f_{\phi}}=\frac{\Omega_f}{\left( \phi+1\right)}\equiv \frac{8\pi G}{3\left( \phi+1\right) }\rho_f$. Substituting  the Eq.(\ref{1e}) into $\Omega_{f_{\phi}}$ we obtain:
\begin{eqnarray}\label{3g}
\Omega_{f_{\phi}}=\frac{c^2}{ H^2L^2}.
\end{eqnarray}
One way to get  an appropriate definition  from $L$, is that we specify it by the future event horizon  for \cite{2c,45}
\begin{eqnarray}\label{4g}
L\equiv d_h(a)=a\int_{t}^{\infty}\frac{dt'}{a(t')}=a\int_{a}^{\infty}\frac{da'}{Ha'^2}.
\end{eqnarray}
Henceforth, we will use $\ln a$, as an independent variable. Therefore, we define, $\dot{X}=\frac{dX}{dt}$,  and  $X'=\frac{dX}{d\ln a}$, so that $\dot{X}=X'H$. By differentiation with respect to coordinate time  from Eq. (\ref{3g}) and that into (\ref{4g}) we have $\dot{d_h}=Hd_h-1$, one can achieve:
   \begin{eqnarray}\label{5g}
\frac{\Omega'_{f_{\phi}}}{\Omega_{f_ {\phi}}}=-\left(\frac{2\dot{H}}{H^2}+2\left( 1-\frac{\sqrt{\Omega_{f_{\phi}} }}{c} \right)    \right).
\end{eqnarray}
In order to clarify  the effect of variety G on the $\Omega_{f_{\phi}}$, we should to get rid of $\dot{H}$ into Eq. (\ref{5g}) in favor of the $\Omega_{f_{\phi}}$. In this regard, differentiation of the Friedmann equation give rise to
\begin{equation}\label{6g}
\dot{H}= \frac{4\pi}{3\left(\phi+1 \right) }\left( G'-3G\left(1+\omega \right) -\frac{G{\phi}'}{\left( \phi+1\right) }\right)\rho,
\end{equation}
where, we using from the fluid equation,  $\dot{\rho}=-3H(1+\omega)\rho$. Since, $\omega$ is
\begin{equation}\label{7g}
\omega=\frac{\omega_{f}\rho_{f}}{\rho}=\frac{\omega_{f}\Omega_{f_{\phi}}}{\Omega_{m_{\phi}} +\Omega_{f_ {\phi}}}=\omega_{f}\Omega_{f_{\phi}},
\end{equation}
 where, according to (\ref{11h}), we using $\Omega_{f_{\phi}}+\Omega_{m_{\phi}}=1$. Now we must obtain the equation of state
  parameter for $\rho_f$ and corresponding pressure $p_f$, $p_f = \omega_f \rho_f$. For this goal, one can    differentiating from  $\rho_f=\frac{3c^2\left(1+\phi \right) }{8\pi G L^2}$ and arrive at
   \begin{equation}\label{g8}
\dot{\rho}_f=\rho_f\left[ \frac{\dot{\phi}}{1+\phi}-2H\left( 1-\frac{1}{LH}\right) \right],
\end{equation}
then, by making  use of the fluid equation $\dot{\rho}_f=-3H\left(1+\omega_f \right)\rho_f $,  we obtain
  \begin{equation}\label{8g}
\omega_{f}=-\frac{1}{3}\left( 1+\frac{2\sqrt{\Omega_{f_{\phi}}}}{c}\right)-\frac{\dot{\phi}}{3H\left(1+\phi \right) }.
\end{equation}
It is remarkable for $\dot{\phi} =0$,  equation (\ref{8g}) reduces  to the $\omega_\Lambda$  of HDE  which is obtained in \cite{45}.
Substituting  Eqs. (\ref{7g},\ref{8g}) into  Eq.(\ref{6g}) gives
\begin{eqnarray}\label{9g}
\frac{2\dot{H}}{H^2}=\left[ {\cal G}-3-\frac{\phi'}{1+\phi}\left( 1-\Omega_{f_{\phi}}\right) + \Omega_{f_{\phi}}+2\Omega^{\frac{3}{2}}_{f_{\phi}}\right]  ,
\end{eqnarray}
where  ${\cal G}$ is $G'/G$. Then by substituting (\ref{9g}) into (\ref{5g}), we have
\begin{eqnarray}\label{10g}
\frac{\Omega'_{f_{\phi}}}{\Omega_{f_ {\phi}}}=-\left[ {\cal G}-1+\Omega_{f_{\phi}}+\left( \Omega_{f_{\phi}}-1\right)\left( \frac{2\sqrt{\Omega_{f_{\phi}}}}{c}+\frac{\phi'}{1+\phi}\right)  \right]
\end{eqnarray}

\section{Some Cosmology Application}
Here, we should follow up an expression relate to the equation of state parameter-like  at the present time.
As this  regards, we have derived the representation for $\Omega'_{f}$, hence  we can found the  $\omega(z)$ form, in the small red shifts. As a rule, in the cosmology literature the same as \cite{46}, one can measure $\omega$, due to $\rho_{f}$ as $\rho_f\sim a^{-3(1+\omega)}$. Expanding $\rho_{f}$  we have:
  \begin{equation}\label{1s}
  \ln \rho_{f}=\ln \rho^0_{f}+ \frac{d\ln \rho_{f}}{d \ln a}\ln a
+\frac{1}{2}\frac{d^2\ln \rho_{f}}{d{(\ln a)}^2}{(\ln a)}^2+\cdots ,
  \end{equation}
here, the derivatives are taken at the present time $a_0=1$. Then, $w(z)$ is given in the small  redshifts $\ln a=-\ln (1+z)=-z$ up to second order,  as:
 \begin{equation}\label{2s}
\omega(z)=-1-\frac{1}{3}\left( \frac{d\ln \rho_{f}}{d \ln a}
-\frac{1}{2}\frac{d^2\ln \rho_{f}}{d{(\ln a)}^2}{(z)}\right).
\end{equation}
Succinctly, we can rewrite (\ref{2s}) as:
\begin{eqnarray}\label{4s}
\omega=\omega_0+\omega_1 z.
\end{eqnarray}
 We replace :
\begin{equation}\label{5s}
\rho_{f}=\frac{3\left(\phi+1 \right)H^2\Omega_{f_{\phi}} }{8\pi G}=\frac{\Omega_{f _{\phi}}\rho_m}{\Omega_{m_{\phi}}}=\frac{\rho_{m_0}\Omega_{f_{\phi}}a^{-3}}{1-\Omega_{f_ {\phi}}},
\end{equation}
At last, calculating the derivatives,   and  some simplification we achieve to $\omega_0,  \omega_1$ as follows:
\begin{equation}\label{6s}
\omega_0=-\frac{1}{3}\frac{\Omega'_{f_{\phi}}}{\Omega_{f_{\phi}}\left(1-\Omega_{f_{\phi}} \right) },
\end{equation}
\begin{eqnarray}\label{7s}
\omega_1=\frac{1}{6\Omega_{f_{\phi}}\left(1-\Omega_{f_{\phi}} \right) }\left[ \Omega''_{f_{\phi}}+\frac{\Omega'^2_{f_{\phi}}\left( 2\Omega_{f_{\phi}}-1\right) }{\Omega_{f_{\phi}}\left(1-\Omega_{f_{\phi}} \right)}
\right].
\end{eqnarray}
 Now, substituting  Eq.(\ref{10g}) into (\ref{6s},\ref{7s}) we obtain:
 \begin{eqnarray}\label{8s}
\omega_o=\frac{1}{3\left(1- \Omega_{f_{\phi}} \right) }\left[ {\cal G}-1+\Omega_{f_{\phi}}+\left( \Omega_{f_{\phi}}-1\right)\left( \frac{2\sqrt{\Omega_{f_{\phi}}}}{c}+\frac{\phi'}{1+\phi}\right) \right],
\end{eqnarray}
  \begin{eqnarray}\label{9s}
\omega_1=\frac{1}{6\left(1-\Omega_{f_{\phi}}\right) }\left[ \frac{{\chi}^2\left( 2\Omega_{f_{\phi}-1} \right) }{1-\Omega_{f_{\phi}}}+\xi\right],
\end{eqnarray}
where, $\chi,\eta,\xi$ are define as follows:
\begin{eqnarray}\label{10s}
\chi&=&{\cal G}-1+\Omega_{f_{\phi}}+\left( \Omega_{f_{\phi}}-1\right)\left( \frac{2\sqrt{\Omega_{f_{\phi}}}}{c}+\frac{\phi'}{1+\phi}\right) ,\cr
\eta&=&\chi+\Omega_{f_{\phi}}+\Omega_{f_{\phi}}\left(\frac{3\sqrt{\Omega_{f_{\phi}}}}{c}+\frac{\phi'}{1+\phi} \right)-\frac{\sqrt{\Omega_{f_{\phi}}}}{c} \cr
\xi&=&\chi\eta-{\cal G'}-\left(\frac{\phi''}{1+\phi}-\frac{{\phi'}^2}{\left(1+\phi \right)^2 }\right) \left( \Omega_{f_{\phi}}-1\right)   .
\end{eqnarray}
In \cite{47} we have obtained  $\phi = \frac{dh(R)}{dR}=\alpha \ln t$
and, $H = \gamma /t$, in which $t$ is the cosmic time. As a result,  the equations (\ref{8s}, \ref{10s})
in terms of  time achieved as
\begin{eqnarray}\label{11s}
\omega_0=\frac{1}{3\left(1- \Omega_{f_{\phi}} \right) }\left[ {\cal G}-1+\Omega_{f_{\phi}}+\left( \Omega_{f_{\phi}}-1\right)\left( \frac{2\sqrt{\Omega_{f_{\phi}}}}{c}+\frac{\alpha}{\gamma\left( 1+\alpha\ln t\right) }\right) \right],\\
\end{eqnarray}
where, $\chi,\vartheta $ and $\xi$ are as
\begin{eqnarray}\label{12s}
\chi&=&{\cal G}-1+\Omega_{f_{\phi}}+\left( \Omega_{f_{\phi}}-1\right)\left( \frac{2\sqrt{\Omega_{f_{\phi}}}}{c}+\frac{\alpha}{\gamma\left( 1+\alpha\ln t\right) }\right) ,\cr
\eta&=&\chi+\Omega_{f_{\phi}}+\Omega_{f_{\phi}}\left(\frac{3\sqrt{\Omega_{f_{\phi}}}}{c}+\frac{\alpha}{\gamma\left( 1+\alpha\ln t\right) } \right)-\frac{\sqrt{\Omega_{f_{\phi}}}}{c},\cr
\xi&=&\chi\eta-{\cal G'}+\frac{{\alpha}^2}{{\gamma}^2\left(1+\alpha\ln t\right)^2 }\left( \Omega_{f_{\phi}}-1\right).
\end{eqnarray}

\section{Conclusion}
Astrophysical observations imply that the  state parameter of dark energy must be changeable on the cosmic time. In this regard, one obvious contender is the holographic dark energy. In the HDE, the parameter $c$
can take  the various values,  but we set $c=1$. However, in comparison with  the holographic dark energy in the $f(R)$,      we have discussed  the holographic dark energy-like whit  the  time dependent gravitation constant  in the framework of $f(R)$ gravity. In the Eq. (\ref{8s}, \ref{9s}), it is considerable that anyone from values $t=0,\infty$  gives a suitable estimate from the state parameter, respectively, as follows.
\begin{eqnarray}\label{1c}
\omega_f(z)=-0.87 +0.074 z,
\end{eqnarray}
For evaluating $\omega_f$ at all the  times we need the observational data for calculating the constants $\alpha$ and $\gamma$.
 Hence, we think that  the modified gravity get a  suitable estimate from $\omega$ at the different times, that is  consistent with  the present local gravity experiments.The last statement is that, however
effect the variation $G$ is very small, but it has an important role  in the evolution of cosmological.

\end{document}